\documentclass{article}
\usepackage{amssymb}
\usepackage{amsmath}

\input{tcilatex}
\begin{document}

\title{Consistent quantization of massless fields of any spin and \\
the generalized Maxwell's equations}
\author{Alexander Gersten and Amnon Moalem \\
Department of Physics\\
Ben Gurion University of the Negev\\
Beer-Sheva, Israel}
\date{Nov.30, 2014}
\maketitle

\begin{abstract}
A simplified formalism of first quantized massless fields of any spin is
presented. The angular momentum basis for particles of zero mass and finite
spin \ $s$\ of the $D^{\left( s-1/2,1/2\right) }$ representation of the
Lorentz group is used to describe the wavefunctions. The advantage of the
formalism is that by equating to zero the \ $s-1$ components of the
wavefunctions, the $2s-1$ subsidiary conditions (needed to eliminate the
non-forward and non-backward helicities) are automatically satisfied.
Probability currents and Lagrangians are derived \ allowing a first
quantized formalism. A simple procedure is derived for connecting the
wavefunctions with potentials and gauge conditions. The spin 1 case is of
particular interest and is described with the $D^{\left( 1/2,1/2\right) }$
vector representation of the well known self-dual representation of the
Maxwell's equations. This representation allows us to generalize Maxwell's
equations by adding the $E_{0}$ and $B_{0}$ components to the electric and
magnetic four-vectors. \ Restrictions on their existence are discussed

\end{abstract}

\section{Introduction}

In a previous paper $\cite{GM12}$ consistent quantization of massless fields
of any spin was developed from first principles. Probability currents and
Lagrangians were derived \ allowing a first quantized formalism. A simple
procedure was derived for connecting the wavefunctions with potentials and
gauge conditions. Here we will present a simplified version of this
formalism and elaborate an extension of the self-dual representation of
Maxwell's equations. This representation allows us to generalize Maxwell's
equations by adding non-zero time like components $E_{0}$ and $B_{0}$ to the
electric and magnetic four-vectors. \ The existence of these components
allows the formation of scalar electromagnetic waves. We will examine some
results.

In our previous paper $\cite{GM12}$ we started our formalism using Dirac's $%
^{\cite{dirac}}$ derivation of equations for massless particles with spin $s$%
, which in the ordinary vector notation are,

\begin{equation}
\frac{1}{s}\left[ s\hat{p}_{0}I^{\left( 2s+1\right) }+S_{x}\hat{p}_{x}+S_{y}%
\hat{p}_{y}+S_{z}\hat{p}_{z}\right] \psi =\left[ \frac{\hat{E}}{c}I^{\left(
2s+1\right) }+\frac{\boldsymbol{S}}{s}\cdot \boldsymbol{\hat{p}}\right] \psi
=0,  \label{a1}
\end{equation}

\begin{equation}
\left[ s\hat{p}_{x}I^{\left( s\right) }+S_{x}\hat{p}_{0}-iS_{y}\hat{p}%
_{z}+iS_{z}\hat{p}_{y}\right] \psi =0,  \label{a2}
\end{equation}

\begin{equation}
\left[ s\hat{p}_{y}I^{\left( s\right) }+S_{y}\hat{p}_{0}-iS_{z}\hat{p}%
_{x}+iS_{x}\hat{p}_{z}\right] \psi =0,  \label{a3}
\end{equation}

\begin{equation}
\left[ s\hat{p}_{z}I^{\left( s\right) }+S_{z}\hat{p}_{0}-iS_{x}\hat{p}%
_{y}+iS_{y}\hat{p}_{x}\right] \psi =0,  \label{a4}
\end{equation}%
where $\psi $ is a $\left( 2s+1\right) $ component wave function and $S_{n}$
are the $\left( 2s+1\right) \times \left( 2s+1\right) $ spin matrices which
satisfy,

\begin{equation}
\left[ S_{x},S_{y}\right] =iS_{z},\quad \left[ S_{z},S_{x}\right]
=iS_{y},\quad \left[ S_{y},S_{z}\right] =iS_{x},\quad
S_{x}^{2}+S_{y}^{2}+S_{z}^{2}=s(s+1)I^{\left( 2s+1\right) }.  \label{a5}
\end{equation}%
In the above, the $\hat{p}_{n}$ are the momenta, $\hat{p}_{0}=\hat{E}/c$, $%
\hat{E}$ the energy, and $I^{\left( 2s+1\right) }$ is a $\left( 2s+1\right)
\times \left( 2s+1\right) $ unit matrix. Eqs. (\ref{a1}-\ref{a4}) were
analyzed extensively by Bacry$^{\cite{bacry}}$, who derived them using
Wigner's condition $^{\cite{wigner}}$ on the Pauli-Lubanski vector $W^{\mu }$
for massless fields,

\begin{equation}
W^{\mu }=s\hat{p}^{\mu },\quad \mu =x_{0},x,y,z,  \label{a6}
\end{equation}

\begin{equation}
W^{\mu }=\frac{-i}{2}\varepsilon ^{\mu \nu \rho \lambda }S_{\rho \lambda }%
\hat{p}_{\nu },\quad S_{\rho \lambda }=\left( 
\begin{array}{cccc}
0 & S_{x} & S_{y} & S_{z} \\ 
-S_{x} & 0 & -iS_{z} & iS_{y} \\ 
-S_{y} & iS_{z} & 0 & -iS_{x} \\ 
-S_{z} & -iS_{y} & iS_{x} & 0%
\end{array}%
\right) \equiv (\mathbf{S,}i\mathbf{S).}  \label{g0}
\end{equation}

Dirac suggested using Eq. (\ref{a1}) as the basic helicity equation and
substitute from it the $\hat{p}_{0}\psi $ into the other 3 equations (\ref%
{a2}-\ref{a4}). Free massless particles may have only two helicity
projections (forward and backward along the momentum vector). Using the
Dirac procedure one obtains the $2s+1$ component Eq. (\ref{a1}) and $2s-1$
independent subsidiary conditions which reduce the number of helicity
projections to 2,

\begin{equation}
\left[ \hat{E}I^{\left( 2s+1\right) }+\frac{c}{s}\boldsymbol{S}\cdot 
\boldsymbol{\hat{p}}\right] \psi ^{\left( 2s+1\right) }=0,  \label{a7}
\end{equation}

\begin{equation}
\left( \mathbf{\Pi \cdot \hat{p}}\right) \psi ^{\left( 2s+1\right) }=0,
\label{a7a}
\end{equation}%
where the components of $\mathbf{\Pi }$ are $\left( 2s-1\right) \times
\left( 2s+1\right) $ matrices of the subsidiary conditions. These two sets
of equations can be combined into $\left( 2s+1\right) +\left( 2s-1\right)
=4s $ equations$^{\cite{GM}}$,$\ $

\begin{equation}
\left[ \hat{E}I^{\left( 4s\right) }+c\boldsymbol{\Gamma }^{\left( 4s\right)
}\cdot \boldsymbol{\hat{p}}\right] \Phi ^{\left( 4s\right) }=0,  \label{a9}
\end{equation}%
equivalent to the former equations, provided that the rows of the $\mathbf{%
\Pi }$ matrices are normalized so that the eigenvalues of $\boldsymbol{%
\Gamma }^{\left( 4s\right) }\cdot \boldsymbol{\hat{p}}$ are $\pm p$ (two
helicities) and,

\begin{equation}
\left( \hat{E}I^{\left( 4s\right) }-c\boldsymbol{\Gamma }^{\left( 4s\right)
}\cdot \boldsymbol{\hat{p}}\right) \left( \hat{E}I^{\left( 4s\right) }+c%
\boldsymbol{\Gamma }^{\left( 4s\right) }\cdot \boldsymbol{\hat{p}}\right)
=\left( \hat{E}^{2}-c^{2}\boldsymbol{\hat{p}}^{2}\right) I^{\left( 4s\right)
},  \label{a10}
\end{equation}%
i.e. they factorize the d'Alembertian. The wave function $\Phi ^{\left(
4s\right) }$ is of the form$^{\cite{GM}}$ (with 2s-1 zeros),

\begin{equation}
\Phi ^{\left( 4s\right) }=\left( 
\begin{array}{c}
0 \\ 
\vdots \\ 
0 \\ 
\psi _{2s+1}^{\left( 2s+1\right) } \\ 
\vdots \\ 
\psi _{-2s-1}^{\left( 2s+1\right) }%
\end{array}%
\right) ,  \label{a11}
\end{equation}%
and the components of $\boldsymbol{\Gamma }^{\left( 4s\right) }$ are $%
4s\times 4s$ matrices, which form a (reducible) representation of the
algebra of the Pauli matrices,

\begin{equation}
\sigma _{0}=\left( 
\begin{array}{cc}
1 & 0 \\ 
0 & 1%
\end{array}%
\right) ,\quad \sigma _{x}=\left( 
\begin{array}{cc}
0 & 1 \\ 
1 & 0%
\end{array}%
\right) ,\quad \sigma _{y}=\left( 
\begin{array}{cc}
0 & -i \\ 
i & 0%
\end{array}%
\right) ,\quad \sigma _{z}=\left( 
\begin{array}{cc}
1 & 0 \\ 
0 & -1%
\end{array}%
\right) .\quad  \label{n3}
\end{equation}

\section{Wave equations for free massless fields}

We shall use the following relations$^{\cite{gersten2}}$,

\begin{equation}
W_{\mu }W^{\mu }=-s\left( s+1\right) \hat{p}_{\mu }\hat{p}^{\mu }I^{\left(
s\right) },\quad W_{\mu }\hat{p}^{\mu }=0,  \label{g1}
\end{equation}%
with the aim of obtaining the Wigner condition Eq. (\ref{a6}) and
simultaneously to factorize the d'Alembertian (in the form $\hat{p}_{\mu }%
\hat{p}^{\mu }$). One can easily see that,

\begin{equation}
\left( W_{\mu }-s\hat{p}_{\mu }\right) \left( W^{\mu }+s\hat{p}^{\mu
}\right) =-s\left( 2s+1\right) \hat{p}_{\mu }\hat{p}^{\mu }I^{\left(
s\right) },  \label{g3}
\end{equation}%
or

\begin{equation}
\left( W_{\mu }+s\hat{p}_{\mu }\right) \left( W^{\mu }-s\hat{p}^{\mu
}\right) =-s\left( 2s+1\right) \hat{p}_{\mu }\hat{p}^{\mu }I^{\left(
s\right) }.  \label{g4}
\end{equation}

From Eqs. (\ref{g3}-\ref{g4}) one can see that if $\psi $, the wavefunction,
satisfies,

\begin{equation}
\left( W^{\mu }+s\hat{p}^{\mu }\right) \psi =0,\quad or\quad \left( W^{\mu
}-s\hat{p}^{\mu }\right) \psi =0,\quad \mu =0,1,2,3,  \label{g5}
\end{equation}%
then the basic massless particle requirement,

\begin{equation}
\hat{p}_{\mu }\hat{p}^{\mu }I^{\left( s\right) }\psi =0,  \label{g6}
\end{equation}%
will be satisfied. \bigskip Explicitly we have,

\begin{equation}
W^{0}=W_{0}=-S_{x}\hat{p}_{x}-S_{y}\hat{p}_{y}-S_{z}\hat{p}_{z},  \label{g61}
\end{equation}%
$\allowbreak $

\begin{equation}
-W^{1}=W_{x}=-S_{x}\hat{p}_{0}-iS_{y}\hat{p}_{z}+iS_{z}\hat{p}_{y},
\label{g62}
\end{equation}

\begin{equation}
-W^{2}=W_{y}=-S_{y}\hat{p}_{0}-iS_{z}\hat{p}_{x}+iS_{x}\hat{p}_{z},
\label{g63}
\end{equation}

\begin{equation}
-W^{3}=W_{z}=-S_{z}\hat{p}_{0}-iS_{x}\hat{p}_{y}+iS_{y}\hat{p}_{x}.
\label{g64}
\end{equation}

Now the main problem is to choose the basis of a space on which the above
operators will act. In Ref. $\cite{GM12}$ we have shown that the angular
momentum basis of the $D^{\left( s-1/2,1/2\right) }$ representation of the
Lorentz group is the proper one to use. This basis is sum of the bases of
spins $s$ and spin $s-1$ with

\begin{equation*}
\left( 2s+1\right) +\left( 2s-1\right) =4s\text{ components.}
\end{equation*}%
In this basis the wavefunction is,

\begin{equation}
\Phi ^{\left( 4s\right) }=\left( 
\begin{array}{c}
\psi _{s-1}^{\left( 2s-1\right) } \\ 
\vdots \\ 
\psi _{-s+1}^{\left( 2s-1\right) } \\ 
\psi _{s}^{\left( 2s+1\right) } \\ 
\vdots \\ 
\psi _{-s}^{\left( 2s+1\right) }%
\end{array}%
\right) \implies \left( 
\begin{array}{c}
0 \\ 
\vdots \\ 
0 \\ 
\psi _{s}^{\left( 2s+1\right) } \\ 
\vdots \\ 
\psi _{-s}^{\left( 2s+1\right) }%
\end{array}%
\right) .  \label{a14}
\end{equation}

The advantage of the formalism, as we have shown in Ref. $\cite{GM12}$, is
that by equating to zero the spin\ $\left( s-1\right) $ components of the
wavefunction $\psi _{\mu }^{\left( 2s-1\right) }$, the $2s-1$ subsidiary
conditions (needed to eliminate the non-forward and non-backward helicities,
according to Wigner's analysis $\cite{wigner}$ are automatically satisfied.
So we are left with equations Eq. $\left( \ref{g5}\right) $ with $\psi =\Phi
^{\left( 4s\right) },$ but not all equations are needed, the equations with $%
\mu =1,2,3$ are the equations Eqs. $\left( \ref{a2}-\ref{a4}\right) $ from
which the subsidiary conditions were derived. Now they are not needed and
instead of Eqs. $\left( \ref{g5}\right) $ we have,

\begin{equation}
\left( s\hat{p}^{0}+W^{0}\right) \Phi ^{\left( 4s\right) }=0,\quad or\quad
\left( s\hat{p}^{0}-W^{0}\right) \Phi ^{\left( 4s\right) }=0,  \label{b1}
\end{equation}%
which are the generalized helicity equations having solutions only in the
forward and backward direction. Note that in Eq. $\left( \ref{b1}\right) $ $%
W^{0}$ is a $4s\times 4s$ matrix. The explicit form of Eq. $\left( \ref{b1}%
\right) $ is given in Refs. $\cite{GM},\cite{GM12}$ with the result,

\begin{equation}
\left[ \hat{E}\Gamma _{0}^{\left( 4s\right) }-c\boldsymbol{\Gamma }^{\left(
4s\right) }\cdot \boldsymbol{\hat{p}}\right] \Phi ^{\left( 4s\right)
}=0,\quad or\quad \left[ \hat{E}\Gamma _{0}^{\left( 4s\right) }+c\boldsymbol{%
\Gamma }^{\left( 4s\right) }\cdot \boldsymbol{\hat{p}}\right] \Phi ^{\left(
4s\right) }=0,  \label{b2}
\end{equation}%
where $\Gamma _{0}^{\left( 4s\right) }=I^{\left( 4s\right) }$ is the $%
4s\times 4s$ unit matrix, $\hat{E}=\hat{p}^{0}/c$, and the $\Gamma $
matrices have the form (in the angular momentum basis),

\begin{equation}
\Gamma _{k}^{\left( 4s\right) }=\frac{1}{s}\left( 
\begin{array}{cc}
\left( -1\right) ^{k+s-1}S_{k}^{\left( 2s-1\right) } & \Pi _{k} \\ 
\Pi _{k}^{H} & \left( -1\right) ^{k+s}S_{k}^{\left( 2s+1\right) }%
\end{array}%
\right) ,\quad k=1,2,3,  \label{b3}
\end{equation}%
where $S_{k}^{\left( 2s-1\right) }$ are the spin $(s-1)$ matrices of Eq. $%
\left( \ref{a5}\right) $, $S_{k}^{\left( 2s+1\right) }$ are the spin $(s)$
matrices of Eq. $\left( \ref{a5}\right) $, $\Pi _{k}$ is the $\left(
2s-1\right) \times \left( 2s+1\right) $\ normalized matrix of the subsidiary
conditions and $\Pi _{k}^{H}$ is the Hermitian conjugate of $\Pi _{k}$. The
matrix elements of $\Pi _{k}$ can be evaluated directly from the normalized
subsidiary conditions

\begin{equation}
\psi _{\mu }^{\left( 2s-1\right) }=\sqrt{s\left( 2s+1\right) }%
\tsum\limits_{m_{1},m_{2}}\langle 1m_{1};sm_{2}|(1s)s-1,\mu \rangle
p_{m_{1}}\psi _{m_{2}}^{\left( 2s+1\right) }=0,  \label{b4}
\end{equation}%
where the expansion is in terms of Clebsch-Gordan coefficients and the
factor $\sqrt{s\left( 2s+1\right) }$ comes from the normalization factor of
the r.h.s. of Eq. $\left( \ref{g3}\right) $.

For example, for spin 1, one obtains,

\begin{equation}
\Pi _{x}=\left( 
\begin{array}{ccc}
-1 & 0 & 0%
\end{array}%
\right) ,\quad \Pi _{y}=\left( 
\begin{array}{ccc}
0 & -1 & 0%
\end{array}%
\right) ,\quad \Pi _{z}=\left( 
\begin{array}{ccc}
0 & 0 & -1%
\end{array}%
\right) ,  \label{b4a}
\end{equation}%
In a Cartesian basis, the $S_{k}^{\left( 3\right) }$ are

\begin{equation}
S_{x}=\left( 
\begin{array}{ccc}
0 & 0 & 0 \\ 
0 & 0 & -i \\ 
0 & i & 0%
\end{array}%
\right) ,\quad S_{y}=\left( 
\begin{array}{ccc}
0 & 0 & i \\ 
0 & 0 & 0 \\ 
-i & 0 & 0%
\end{array}%
\right) ,\quad S_{z}=\left( 
\begin{array}{ccc}
0 & -i & 0 \\ 
i & 0 & 0 \\ 
0 & 0 & 0%
\end{array}%
\right) ,  \label{n8}
\end{equation}

\begin{equation}
\Gamma _{k}^{\left( 4\right) }=\left( 
\begin{array}{cc}
0 & \Pi _{k} \\ 
\Pi _{k}^{H} & S_{k}^{\left( 3\right) }%
\end{array}%
\right) ,\quad k=1,2,3,\quad \Gamma _{0}^{\left( 4\right) }=\text{unit
matrix,}  \label{b4b}
\end{equation}

For spin 2, one obtains,

\begin{equation}
\Pi _{x}=\left( 
\begin{array}{ccccc}
\sqrt{3} & 0 & -\frac{1}{\sqrt{2}} & 0 & 0 \\ 
0 & \sqrt{\frac{3}{2}} & 0 & -\sqrt{\frac{3}{2}} & 0 \\ 
0 & 0 & \frac{1}{\sqrt{2}} & 0 & -\sqrt{3}%
\end{array}%
\right) ,  \label{b5}
\end{equation}

\begin{equation}
\Pi _{y}=i\left( 
\begin{array}{ccccc}
-\sqrt{3} & 0 & -\frac{1}{\sqrt{2}} & 0 & 0 \\ 
0 & -\sqrt{\frac{3}{2}} & 0 & -\sqrt{\frac{3}{2}} & 0 \\ 
0 & 0 & -\frac{1}{\sqrt{2}} & 0 & -\sqrt{3}%
\end{array}%
\right) ,  \label{b6}
\end{equation}

\begin{equation}
\Pi _{z}=\left( 
\begin{array}{ccccc}
0 & -\sqrt{3} & 0 & 0 & 0 \\ 
0 & 0 & -2 & 0 & 0 \\ 
0 & 0 & 0 & -\sqrt{3} & 0%
\end{array}%
\right) .  \label{b7}
\end{equation}

\begin{equation}
\Gamma _{k}^{\left( 8\right) }=\frac{1}{2}\left( 
\begin{array}{cc}
\left( -1\right) ^{k+1}S_{k}^{\left( 3\right) } & \Pi _{k} \\ 
\Pi _{k}^{H} & \left( -1\right) ^{k}S_{k}^{\left( 5\right) }%
\end{array}%
\right) ,\quad k=1,2,3,  \label{b10}
\end{equation}

\bigskip More details can be found in Refs. $\cite{GM}-\cite{GM12}.$

\section{\protect\LARGE First quantization}

\bigskip As is in the case of the Klein-Gordon or Dirac equations, we
substitute in Eqs. (\ref{b2})

\begin{equation}
\hat{E}\Longrightarrow i\hbar \frac{\partial }{\partial t},\quad \mathbf{%
\hat{p}}\mathbf{\Longrightarrow -}i\hbar \mathbf{\nabla },\quad
\end{equation}%
with a result

\begin{equation}
i\hbar \left( \Gamma _{0}^{\left( 4s\right) }\frac{\partial }{\partial t}+c%
\boldsymbol{\Gamma }^{\left( 4s\right) }\cdot \mathbf{\nabla }\right) \Phi
^{\left( 4s\right) }=0.  \label{a16}
\end{equation}

Similarly to the Schroedinger and Dirac equations we define the Hamiltonian $%
\mathcal{H}$ from Eq. (\ref{a16}),

\begin{equation}
i\hbar \frac{\partial }{\partial t}\Phi ^{\left( 4s\right) }=-i\hbar c%
\boldsymbol{\Gamma }^{\left( 4s\right) }\cdot \mathbf{\nabla }\Phi ^{\left(
4s\right) }=\mathcal{H}\Phi ^{\left( 4s\right) },  \label{a17}
\end{equation}%
and find a conserved probability current (the superscript $H$ denotes the
Hermitian conjugate),

\begin{equation}
\partial _{t}\left( \left( \Phi ^{\left( 4s\right) }\right) ^{H}\Phi
^{\left( 4s\right) }\right) +\frac{c}{s}\mathbf{\nabla }\mathbf{\cdot }%
\left( \left( \Phi ^{\left( 4s\right) }\right) ^{H}\mathbf{\tilde{S}}\Phi
^{\left( 4s\right) }\right) =0.  \label{a18}
\end{equation}%
Thus the probability density (which should be normalized) is, 
\begin{equation}
\rho =\left( \Phi ^{\left( 4s\right) }\right) ^{H}\Phi ^{\left( 4s\right)
},\quad \int \int \int dxdydz\rho =1.  \label{a18a}
\end{equation}%
In the above the density is non-negative and the scalar product is positive
definite. Having this result we can find a Lagrangian density,

\begin{equation}
\mathcal{L=}i\hbar \left( \Phi ^{\left( 4s\right) }\right) ^{H}\left( \Gamma
_{0}^{4s}\frac{\partial }{\partial t}+c\boldsymbol{\Gamma }^{\left(
4s\right) }\cdot \mathbf{\nabla }\right) \Phi ^{\left( 4s\right) },
\label{a19}
\end{equation}%
and using the definition of the energy momentum tensor $T^{\mu \nu }$ , we
find$^{\cite{GM}}$,

\begin{center}
\begin{equation}
\int \int \int dxdydzT^{00}=\int \int \int dxdydz\left( \Phi ^{\left(
4s\right) }\right) ^{H}\mathcal{H}\Phi ^{\left( 4s\right) }=<\mathcal{H}>,
\label{a20}
\end{equation}%
$\qquad $
\end{center}

\begin{equation}
\int \int \int dxdydzT^{0k}=\int \int \int dxdydz\left( \Phi ^{\left(
4s\right) }\right) ^{H}\left( c\hat{p}_{k}\right) \Phi ^{\left( 4s\right)
}=<c\hat{p}_{k}>,\quad k=1,2,3,  \label{a21}
\end{equation}%
i.e. consistent with the expectation values of energy and momentum.

\section{\protect\bigskip Wave equations for free fields}

From the Lagrangian Eq. (\ref{a19}) we obtain the two equations Eqs. (\ref%
{b2}), for particles of spin $s$ (4$s$ is the dimension of the matrices),%
\begin{equation}
i\hbar \left( \Gamma _{0}^{\left( 4s\right) }\frac{\partial }{\partial t}+c%
\boldsymbol{\Gamma }^{\left( 4s\right) }\cdot \mathbf{\nabla }\right) \Phi
^{\left( 4s\right) }=0,  \label{b8}
\end{equation}

\begin{equation}
i\hbar \left( \Gamma _{0}^{\left( 4s\right) }\frac{\partial }{\partial t}-c%
\boldsymbol{\Gamma }^{\left( 4s\right) }\cdot \mathbf{\nabla }\right) \Phi
^{\left( 4s\right) }=0.  \label{b9}
\end{equation}

The $\Gamma $\ matrices form a reducible representation of the Pauli
matrices Eq. (\ref{n3}). For the spin 
$\frac12$
Eqs. ( \ref{b8}-\ref{b9}) coincide with the massless neutrino equations. The 
$\Gamma $\ matrices have the same eigenvalues as the Pauli matrices $\pm 1$
(2$s$ degenerate). With the built-in subsidiary conditions only two
solutions of each of the equations Eqs. ( \ref{b8}-\ref{b9}) are possible.
Thus Eqs. ( \ref{b8}-\ref{b9}) are the restricted to forward or backward
helicity equations for free massless particles. As the energy operator is
proportional to the helicity the two solutions only differ in sign of the
energy, i.e. one solution corresponds to positive energy and the second to
negative energy. One can also note that the solutions of Eq. (\ref{b8}) are
also solutions of Eq. (\ref{b9}) with the opposite sign of the helicity.
Therefore Eq. (\ref{b8}) has positive energy solution in the helicity's
forward direction; its negative energy solution in the helicity's backward
direction is the positive energy solution of Eq. (\ref{b9}) for the
helicity's backward direction.

In Summary Eq. (\ref{b8}) is the equation of the helicity in the forward
direction and Eq. (\ref{b9}) is the equation for the helicity in the
backward direction, if the physical requirement of positive energy is
imposed.

\subsubsection{Examples}

For spin 1, the photon with helicity $\pm 1$ and total momentum $P=\sqrt{%
p_{x}^{2}+p_{y}^{2}+p_{z}^{2}}$, the normalized solution of Eq. (\ref{b8})
(in Cartesian basis) for positive energy and forward helicity is

\begin{equation}
\Phi ^{\left( 4\right) }=\frac{1}{\sqrt{2}P}\allowbreak \allowbreak \left( 
\begin{array}{c}
0 \\ 
-\frac{p_{x}p_{z}-ip_{y}P}{\sqrt{p_{x}^{2}+p_{y}^{2}}} \\ 
-\frac{ip_{x}P+p_{y}p_{z}}{\sqrt{p_{x}^{2}+p_{y}^{2}}} \\ 
\sqrt{p_{x}^{2}+p_{y}^{2}}%
\end{array}%
\right) \exp (\frac{1}{i\hbar }(Et-\boldsymbol{p\cdot x)),}  \label{b11}
\end{equation}

and the normalized solution of Eq. (\ref{b9}) (in Cartesian basis) for
positive energy and backward helicity is

\begin{equation}
\Phi ^{\left( 4\right) }=\frac{1}{\sqrt{2}P}\left( 
\begin{array}{c}
0 \\ 
-\frac{ip_{y}P+p_{x}p_{z}}{\sqrt{p_{x}^{2}+p_{y}^{2}}} \\ 
-\frac{p_{y}p_{z}-ip_{x}P}{\sqrt{p_{x}^{2}+p_{y}^{2}}} \\ 
\sqrt{p_{x}^{2}+p_{y}^{2}}%
\end{array}%
\right) \exp (\frac{-1}{i\hbar }(Et-\boldsymbol{p\cdot x)).}  \label{b12}
\end{equation}

For spin 2, the graviton with helicity $\pm 2$ and total momentum $P=\sqrt{%
p_{x}^{2}+p_{y}^{2}+p_{z}^{2}}$, the normalized solution of Eq. (\ref{b8})
(in angular momentum basis) for positive energy and forward helicity is

$\allowbreak $%
\begin{equation}
\Phi ^{\left( 8\right) }=\frac{\left( p_{x}^{2}+p_{y}^{2}\right) ^{2}}{4P^{4}%
}\left( 
\begin{array}{c}
0 \\ 
0 \\ 
0 \\ 
\frac{1}{2}\left( \frac{p_{z}-P}{p_{x}-ip_{y}}\right) ^{2} \\ 
\left( \frac{p_{z}-P}{p_{x}-ip_{y}}\right) \\ 
\sqrt{\frac{3}{2}} \\ 
\frac{p_{x}-ip_{y}}{p_{z}-P} \\ 
\frac{1}{2}\left( \frac{p_{x}-ip_{y}}{p_{z}-P}\right) ^{2}%
\end{array}%
\right) \exp (\frac{1}{i\hbar }(Et-\boldsymbol{p\cdot x)),},  \label{b13}
\end{equation}

and the normalized solution of Eq. (\ref{b9}) (in angular momentum basis)
for positive energy and backward helicity is

\begin{equation}
\Phi ^{\left( 8\right) }=\frac{\left( p_{x}^{2}+p_{y}^{2}\right) ^{2}}{4P^{4}%
}\left( 
\begin{array}{c}
0 \\ 
0 \\ 
0 \\ 
\frac{1}{2}\left( \frac{p_{z}+P}{p_{x}-ip_{y}}\right) ^{2} \\ 
\left( \frac{p_{z}+P}{p_{x}-ip_{y}}\right) \\ 
\sqrt{\frac{3}{2}} \\ 
\frac{p_{x}-ip_{y}}{p_{z}+P} \\ 
\frac{1}{2}\left( \frac{p_{x}-ip_{y}}{p_{z}+P}\right) ^{2}%
\end{array}%
\right) \exp (\frac{-1}{i\hbar }(Et-\boldsymbol{p\cdot x)).}  \label{b14}
\end{equation}

\section{Generalized Maxwell's equations}

\subsection{Maxwell's equations}

Maxwell's equations $\cite{Jackson}$\ in Gaussian system units are,

\begin{equation}
\mathbf{\nabla }\cdot \mathbf{E}=\alpha j_{0},  \label{b15}
\end{equation}

\begin{equation}
\mathbf{\nabla }\times \mathbf{B-}\partial _{0}\boldsymbol{E}=\alpha 
\boldsymbol{j},  \label{b16}
\end{equation}

\begin{equation}
\mathbf{\nabla }\cdot \mathbf{B}=0,  \label{b17}
\end{equation}

\begin{equation}
\mathbf{\nabla }\times \mathbf{E+}\partial _{0}\boldsymbol{B}=0,  \label{b18}
\end{equation}%
where $\alpha =\frac{4\pi }{c}\allowbreak ,\quad j_{0}=c\rho ,\quad \partial
_{0}=\frac{1}{c}\frac{\partial }{\partial t}.$

Minkowski$\left[ \cite{minkowski}\right] $ found the covariant form of
Maxwell's equations,

\begin{equation}
\partial _{\mu }F^{\mu \nu }=\alpha j^{\nu },\quad \partial _{\mu }\tilde{F}%
^{\mu \nu }=0,\quad \nu =0,1,2,3,  \label{40}
\end{equation}%
where the antisymmetric tensor $F^{\mu \nu }$ and its dual $\tilde{F}^{\mu
\nu }$ are defined via the electric and magnetic fields $\mathbf{E}\mathbf{\ 
}$and $\mathbf{B}$ respectively as,

\begin{equation}
\left( F^{\mu \nu }\right) =\left( 
\begin{array}{cccc}
0 & -E_{x} & -E_{y} & -E_{z} \\ 
E_{x} & 0 & -B_{z} & B_{y} \\ 
E_{y} & B_{z} & 0 & -B_{x} \\ 
E_{z} & -B_{y} & B_{x} & 0%
\end{array}%
\right) ,  \label{40a}
\end{equation}

\begin{equation}
\tilde{F}^{\mu \nu }=\frac{1}{2}\epsilon ^{\mu \nu \alpha \beta }F_{\alpha
\beta }=\left( 
\begin{array}{cccc}
0 & -B_{x} & -B_{y} & -B_{z} \\ 
B_{x} & 0 & E_{z} & -E_{y} \\ 
B_{y} & -E_{z} & 0 & E_{x} \\ 
B_{z} & E_{y} & -E_{x} & 0%
\end{array}%
\right) ,  \label{40c}
\end{equation}%
where $\epsilon ^{\mu \nu \alpha \beta }$ is the totally antisymmetric
tensor ($\epsilon ^{0123}=1,\quad \epsilon ^{\mu \nu \alpha \beta
}=-\epsilon _{\mu \nu \alpha \beta }$). The sum of the $F^{\mu \nu }$ and $i%
\tilde{F}^{\mu \nu }$ is the self-dual antisymmetric tensor, which depends
only on the combination

\begin{equation}
\mathbf{\Psi }=\left( \mathbf{E}\mathbf{+}i\mathbf{B}\right) ,  \label{40d}
\end{equation}

\begin{equation}
F^{\mu \nu }+i\tilde{F}^{\mu \nu }=\left( 
\begin{array}{cccc}
0 & -E_{x}-iB_{x} & -E_{y}-iB_{y} & -E_{z}-iB_{z} \\ 
E_{x}+iB_{x} & 0 & iE_{z}-B_{z} & B_{y}-iE_{y} \\ 
E_{y}+iB_{y} & B_{z}-iE_{z} & 0 & iE_{x}-B_{x} \\ 
E_{z}+iB_{z} & iE_{y}-B_{y} & B_{x}-iE_{x} & 0%
\end{array}%
\right) ,  \label{41}
\end{equation}

$\ \ \ \ \ \ \allowbreak $%
\begin{equation}
=\left( 
\begin{array}{cccc}
0 & -\Psi _{x} & -\Psi _{y} & -\Psi _{z} \\ 
\Psi _{x} & 0 & i\Psi _{z} & -i\Psi _{y} \\ 
\Psi _{y} & -i\Psi _{z} & 0 & i\Psi _{x} \\ 
\Psi _{z} & i\Psi _{y} & -i\Psi _{x} & 0%
\end{array}%
\right)  \label{41a}
\end{equation}%
$.$

The self dual Maxwell's equations take the form $\cite{GM}$,

\begin{equation}
\partial _{\mu }\left( F^{\mu \nu }+i\tilde{F}^{\mu \nu }\right) =-\left(
\Gamma _{\mu }^{\left( 4\right) }\right) ^{\nu i}\partial _{\mu }\Psi
_{i}=\alpha j^{\nu },  \label{43}
\end{equation}%
where the $\Gamma _{\mu }^{\left( 4\right) }$ matrices are the spin 1
matrices defined by Eq. (\ref{b4b}). Thus within the $D^{\left( 
{\frac12}%
,%
{\frac12}%
\right) }$ representation of the Lorentz group, Maxwell's equation can be
written as,

\begin{equation}
\left( \Gamma _{0}^{\left( 4\right) }\partial _{0}+\boldsymbol{\Gamma }%
^{\left( 4\right) }\cdot \mathbf{\nabla }\right) \Phi ^{\left( 4\right)
}=-\alpha J,\quad \Phi ^{\left( 4\right) }=\left( 
\begin{array}{c}
0 \\ 
\Psi _{x} \\ 
\Psi _{y} \\ 
\Psi _{z}%
\end{array}%
\right) ,\quad J=\left( 
\begin{array}{c}
c\rho \\ 
j_{x} \\ 
j_{y} \\ 
j_{z}%
\end{array}%
\right) .  \label{44}
\end{equation}

\bigskip Eq. (\ref{44}) is the same as Eq. (\ref{43}) which is Lorentz
covariant, thus Eq. (\ref{44}) is also Lorentz invariant. This invariance
also applies to the zero component of $\Phi ^{\left( 4\right) }$, which was
proven explicitly by Lomont $\cite{lomont}.$ For $J=0$, the free Maxwell
equations takes the form,

\begin{equation}
\left( \Gamma _{0}^{\left( 4\right) }\partial _{0}+\boldsymbol{\Gamma }%
^{\left( 4\right) }\cdot \mathbf{\nabla }\right) \Phi ^{\left( 4\right) }=0,
\label{45}
\end{equation}

which is exactly Eq. (\ref{b8}). The zero component is the perpendicularity
condition for the photon equation and it ensures that the helicity can be
only in the forward or backward direction. It also insures that the photon
will be a particle of spin 1. This condition also remains for the
inhomogenious Maxwell's equations Eq. (\ref{44}). But in the presence of
interactions it is not obvious that the spin 1 property should remain.
Therefore we will now generalize the non-homogenious Maxwell equations and
replace the zero component with $\Psi _{0}=E_{0}+iB_{0}.$

\subsection{The generalized equations}

The generalized equations are,

\begin{equation}
\left( \Gamma _{0}^{\left( 4\right) }\partial _{0}+\boldsymbol{\Gamma }%
^{\left( 4\right) }\cdot \mathbf{\nabla }\right) \Phi ^{\left( 4\right)
}=-\alpha J,\quad \Phi ^{\left( 4\right) }=\left( 
\begin{array}{c}
\Psi _{0} \\ 
\Psi _{x} \\ 
\Psi _{y} \\ 
\Psi _{z}%
\end{array}%
\right) ,\quad J=\left( 
\begin{array}{c}
c\rho \\ 
j_{x} \\ 
j_{y} \\ 
j_{z}%
\end{array}%
\right) .  \label{46}
\end{equation}%
from which one can extract the generalized Maxwell's equations in the form

\begin{equation}
\partial _{0}E_{0}+\boldsymbol{\nabla }\cdot \boldsymbol{E}=\alpha j_{0},
\label{65}
\end{equation}

\begin{equation}
\partial _{0}\boldsymbol{E+}\mathbf{\nabla }E_{0}-\mathbf{\nabla \times B}%
=-\alpha \boldsymbol{j},  \label{66}
\end{equation}

\begin{equation}
\partial _{0}B_{0}+\nabla \cdot \boldsymbol{B}=0,  \label{67}
\end{equation}

\begin{equation}
\boldsymbol{\nabla }\times \boldsymbol{E+\partial }_{0}\boldsymbol{B+\nabla }%
B_{0}=0.  \label{68}
\end{equation}

Similar equations were derived by Dvoeglazov$\cite{dvoeglazov}.$ \ We will
now look for new features of these equations.

\subsection{Energy density}

According to Eqs. (\ref{a17}) and (\ref{a20}) the energy density is
proportional to,

\begin{equation}
\left( 
\begin{array}{cccc}
\Psi _{0}^{\ast } & \Psi _{x}^{\ast } & \Psi _{y}^{\ast } & \Psi _{z}^{\ast }%
\end{array}%
\right) \left( i\hbar c\partial _{0}\right) \left( 
\begin{array}{c}
\Psi _{0} \\ 
\Psi _{x} \\ 
\Psi _{y} \\ 
\Psi _{z}%
\end{array}%
\right) =i\hbar c\left( \Psi _{0}^{\ast }\partial _{0}\Psi _{0}+\mathbf{\Psi 
}^{\ast }\mathbf{\cdot }\left( \partial _{0}\mathbf{\Psi }\right) \right)
\label{46a}
\end{equation}

\begin{equation}
=i\hbar c\left[ \left( E_{0}-iB_{0}\right) \partial _{0}\left(
E_{0}+iB_{0}\right) +\left( \mathbf{E}-i\mathbf{B}\right) \mathbf{\cdot }%
\left( \partial _{0}\left( \mathbf{E}+i\mathbf{B}\right) \right) \right] ,
\label{46b}
\end{equation}%
which is different from the conventional form proportional to

\begin{equation*}
\mathbf{\Psi }^{\ast }\mathbf{\cdot \Psi }=\left( \mathbf{E}-i\mathbf{B}%
\right) \mathbf{\cdot }\left( \mathbf{E}+i\mathbf{B}\right) =\mathbf{E}^{2}+%
\mathbf{B}^{2}
\end{equation*}

\subsection{The potentials}

Let us introduce the potentials,

\begin{equation}
A^{\left( 4\right) }=\left( 
\begin{array}{c}
A_{0} \\ 
A_{x} \\ 
A_{y} \\ 
A_{z}%
\end{array}%
\right) ,  \label{47}
\end{equation}%
where the scalar potential $V=cA_{0}$.\ \ Their relation to the wavefunction 
$\Phi ^{(4)}$ can be obtained by requiring that in the Lorenz gauge they
satisfy the wave equation,

\begin{equation}
\partial _{\mu }\partial ^{\mu }A^{(4)}=\alpha J.  \label{48}
\end{equation}%
The $\Gamma $ matrices, which are a representation of the Pauli matrices,
will factorize the d'Alembertian the same way as the Pauli matrices
factorize it, i.e., 
\begin{equation}
\left( \Gamma _{0}^{\left( 4\right) }\partial _{0}-\boldsymbol{\Gamma }%
^{\left( 4\right) }\cdot \mathbf{\nabla }\right) \left( \Gamma _{0}^{\left(
4\right) }\partial _{0}+\boldsymbol{\Gamma }^{\left( 4\right) }\cdot \mathbf{%
\nabla }\right) =\Gamma _{0}^{\left( 4\right) }\partial _{\mu }\partial
^{\mu }  \label{49}
\end{equation}%
In order to get Eq. (\ref{48}), using Eqs. (\ref{46}) and (\ref{49}) the
relation between the $\Phi ^{(4)}$ and $A^{\left( 4\right) }$ has to be

\begin{equation}
\Phi ^{\left( 4\right) }=\left( 
\begin{array}{c}
E_{0}+iB_{0} \\ 
E_{x}+iB_{x} \\ 
E_{y}+iB_{y} \\ 
E_{z}+iB_{z}%
\end{array}%
\right) =-\left( \Gamma _{0}^{\left( 4\right) }\partial _{0}-\boldsymbol{%
\Gamma }^{\left( 4\right) }\cdot \mathbf{\nabla }\right) \left( 
\begin{array}{c}
A_{0} \\ 
A_{x} \\ 
A_{y} \\ 
A_{z}%
\end{array}%
\right)  \label{50}
\end{equation}

\begin{equation}
=-\left( 
\begin{array}{c}
\partial _{0}A_{0}+\partial _{1}A_{x}+\partial _{2}A_{y}+\partial _{3}A_{z}
\\ 
\partial _{0}A_{x}+\partial _{1}A_{0}+i\left( \partial _{2}A_{z}-\partial
_{3}A_{y}\right) \\ 
\partial _{0}A_{y}+\partial _{2}A_{0}-i\left( \partial _{1}A_{z}-\partial
_{3}A_{x}\right) \\ 
\partial _{0}A_{z}+\partial _{3}A_{0}+i\left( \partial _{1}A_{y}-\partial
_{2}A_{x}\right)%
\end{array}%
\right) ,  \label{51}
\end{equation}%
from which the relation between the fields and the potentials (in the Lorenz
gauge) is,

\begin{equation}
E_{0}=-\partial _{0}A_{0}-\nabla \cdot \mathbf{A,}  \label{52}
\end{equation}

\begin{equation}
B_{0}=0,  \label{53}
\end{equation}

\begin{equation}
\mathbf{E=-\nabla }A_{0}-\partial _{0}\mathbf{A,}  \label{54}
\end{equation}

\begin{equation}
\mathbf{B=\nabla \times A.}  \label{55}
\end{equation}

Eqs. (\ref{52}) and (\ref{53}) are new relations of the generalized
Maxwell's equations, but the Lorenz condition,

\begin{equation}
\partial _{0}A_{0}+\nabla \cdot \mathbf{A}=0,  \label{56}
\end{equation}%
enforces the spin 0 component $E_{0}$\ to be zero, this result should be
gauge invariant as the electric and magnetic fields are. Thus only when the
Lorenz condition is not satisfied (while Eq. (\ref{48}) is valid) the $E_{0}$
can be different from zero. This result indicates that spin 1 is conserved
in Maxwell's equations. In order to have $E_{0}\neq 0$ a spin changing
interaction must be added to the 4-current, as the $E_{0}$\ component
belongs to the spin 0 part of the four vector $E_{\mu }$ which belongs to
the $D^{\left( 
{\frac12}%
,%
{\frac12}%
\right) }$\ representation of the Lorentz group with an angular momentum
basis of spin 0 and spin 1.\ \ \ 

\subsection{Wave equations}

Let us multiply Eq. (\ref{46}) by $\left( \Gamma _{0}^{\left( 4\right)
}\partial _{0}-\boldsymbol{\Gamma }^{\left( 4\right) }\cdot \mathbf{\nabla }%
\right) ,$

\begin{equation}
\Gamma _{0}^{\left( 4\right) }\partial _{\mu }\partial ^{\mu }\Phi ^{\left(
4\right) }=\left( \Gamma _{0}^{\left( 4\right) }\partial _{0}-\boldsymbol{%
\Gamma }^{\left( 4\right) }\cdot \mathbf{\nabla }\right) \left( \Gamma
_{0}^{\left( 4\right) }\partial _{0}+\boldsymbol{\Gamma }^{\left( 4\right)
}\cdot \mathbf{\nabla }\right) \Phi ^{\left( 4\right) }=-\alpha \left(
\Gamma _{0}^{\left( 4\right) }\partial _{0}-\boldsymbol{\Gamma }^{\left(
4\right) }\cdot \mathbf{\nabla }\right) J,  \label{57}
\end{equation}%
from which we obtain,

\begin{equation}
\square E_{0}=\alpha \left( \partial _{0}j_{0}+\mathbf{\nabla \cdot j}%
\right) ,  \label{58}
\end{equation}

\begin{equation}
\square B_{0}=0,  \label{59}
\end{equation}

\begin{equation}
\square \mathbf{E=}\alpha \left( \mathbf{\partial }_{0}\mathbf{j+\mathbf{%
\nabla }}j\mathbf{_{0}}\right) ,  \label{60}
\end{equation}

\begin{equation}
\square \mathbf{B=-}\alpha \mathbf{\nabla \times j,}  \label{61}
\end{equation}%
where $\square =\partial _{\mu }\partial ^{\mu }.$

Current conservation requires,

\begin{equation}
\partial _{0}j_{0}+\mathbf{\nabla \cdot j=0,}  \label{62}
\end{equation}%
therefore when the current is conserved

\begin{equation}
\square E_{0}=0,  \label{63}
\end{equation}%
and only if the current is not conserved the possibility of $\square
E_{0}\neq 0$ will exist. As indicated before, a spin changing interaction
can induce the $E_{0}$ component$.$

\subsection{\protect\bigskip Scalar electrodynamic waves}

There are claims that scalar electrodynamic waves have been detected \cite%
{monstein}. As we have shown, within the framework of electrodynamics with
conserved currents the scalar waves cannot exist. There exists a
misconception that longitudinal waves are the scalar waves. Longitudinal
waves can exist as solutions of Maxwell's equations when the
perpendicularity condition $\mathbf{\nabla \cdot E=\nabla \cdot B=0}$ is \
violated, for instance when $\mathbf{\nabla \cdot E=}\alpha j\mathbf{_{0}}$,
i.e. in the presence of a charge distribution. Therefore in Ref. \cite%
{monstein} the longitudinal waves were actually vector waves. Other claims
of scalar electromagnetic waves come from K. Meyl \cite{meyl}, which are
based on a theory different from Maxwell's. A modification of Maxwell's
equation to allow for current non-conservation was proposed by Hively and
Giacos \cite{hively} with the aim to experimentaly detect scalar
electromagnetic waves.

\subsection{Summary and conclusions}

A simplified formalism of first quantized massless fields of any spin is
presented. Angular momentum basis for particles of zero mass and finite spin
\ $s$\ of the $D^{\left( s-1/2,1/2\right) }$ representation of the Lorentz
group is used to describe the wavefunctions. The advantage of the formalism
is that by equating to zero the \ $s-1$ components of the wavefunctions, the 
$2s-1$ subsidiary conditions (needed to eliminate the non-forward and
non-backward helicities) are automatically satisfied. Probability currents
and Lagrangians are derived \ allowing a first quantized formalism. A simple
procedure is derived of connecting the wavefunctions with potentials and
gauge conditions.

The spin 1 case is of particular interest and is described with the $%
D^{\left( 1/2,1/2\right) }$ vector representation of the well known
self-dual representation of the Maxwell's equations Eq. (\ref{44}). The wave
function is the 4-vector

\begin{equation}
\Phi ^{\left( 4\right) }=\left( 
\begin{array}{c}
0 \\ 
\Psi _{x} \\ 
\Psi _{y} \\ 
\Psi _{z}%
\end{array}%
\right) =\left( 
\begin{array}{c}
0 \\ 
E_{x}+iB_{x} \\ 
E_{y}+iB_{y} \\ 
E_{z}+iB_{z}%
\end{array}%
\right) .  \label{64}
\end{equation}

The zero component remains unchanged under Lorentz transformations \cite%
{lomont},\cite{GM}. It is well known that the vector $D^{\left(
1/2,1/2\right) }$ representation has an angular momentum basis of spin 1
plus spin 0. The meening of it is that the wave function Eq. (\ref{64}%
)contains only the spin 1 contribution. Thus Maxwell's equations contain an
intrinsic spin 1. This representation allows to generalize Maxwell's
equations by adding the $E_{0}$ and $B_{0}$ timelike components to create
the electric and magnetic four-vectors, \ 

\begin{equation}
\Phi ^{\left( 4\right) }=\left( 
\begin{array}{c}
E_{0}+iB_{0} \\ 
E_{x}+iB_{x} \\ 
E_{y}+iB_{y} \\ 
E_{z}+iB_{z}%
\end{array}%
\right) .  \label{70}
\end{equation}

We found that as long as the electromagnetic current is conserved, $E_{0}$
and $B_{0}$ have to be zero. Thus scalar electromagnetic waves can be
created only if charged conservation is violated, or if a spin changing
interaction is added to the conserved current.

\end{document}